# Scalable monolayer-functionalized nanointerface for thermal conductivity enhancement in copper/diamond composite

*Bin Xu[1], Shih-Wei Hung[2], Shiqian Hu[1], Cheng Shao[1], Rulei Guo[1], Junho Choi[1], Takashi Kodama[1], Fu-Rong Chen[2], Junichiro Shiomi[1*]*

[1]Department of Mechanical Engineering, The University of Tokyo, Tokyo, Japan

[2]Department of Materials Science and Engineering, City University of Hong Kong, Hong Kong, China

*Corresponding author. Tel: 03-5841-6283 Email: shiomi@photon.t.u-tokyo.ac.jp

**ABSTRACT:** Aiming at developing high thermal conductivity copper/diamond composite, an unconventional approach applying self-assembled monolayer (SAM) prior to the high-temperature sintering of copper/diamond composite was utilized to enhance the thermal boundary conductance (TBC) between copper and diamond. The enhancement was first systematically confirmed on a model interface system by detailed SAM morphology characterization and TBC measurements. TBC significantly depends on the SAM coverage and ordering, and the formation of high-quality SAM promoted the TBC to 73 MW/m$^2$-K from 27 MW/m$^2$-K, the value without SAM. With the help of molecular dynamics simulations, the TBC enhancement was identified to be determined by the number of SAM bridges and the overlap of vibrational density of states. The diamond particles of 210 μm in size were simultaneously functionalized by SAM with the condition giving the highest TBC in the model system and sintered together with the copper to fabricate isotropic copper/diamond composite of 50% volume fraction. The measured thermal conductivity marked 711 W/m-K at room temperature, the highest value among the ones with similar diamond-particles volume fraction and size. This work demonstrates a novel strategy to enhance the thermal conductivity of composite materials by SAM functionalization.

1. INTRODUCTION

With the advance in high-performance power electronic devices and the accompanying increase in heat generation density, there is a growing need for technology that realizes higher heat dissipation. There, the development of materials with a thermal conductivity higher than metals like copper (Cu, 400 W/m-K at room temperature) is essential. A way to realize this without losing the merit of metals, such as surface treatability and mechanical compliance that makes them connectable to the surrounding components, is to form a composite by adding higher thermal conductivity particles. Cu/diamond composite is a promising candidate in this context because of the extremely high intrinsic thermal conductivity of diamond. However, according to the effective medium theory (EMT)[1], thermal boundary conductance (TBC)[2] at the interface between Cu matrix and diamond particles plays a crucial role in determining the thermal conductivity of the composite, especially when the particle size is on the order of micrometers, which is realistic considering the material cost that increases with the size of the diamond particles.

Many studies have been carried out to enhance the TBC of the pristine Cu/diamond interface, and a few have directly probed the TBC using the time-domain thermoreflectance (TDTR) method. Approaches like the precise control over the chemical component[3], pressure[4], and the introduction of carbide buffer layers (like TiC[5][6]) was found effective to enhance the TBC. Among the above approaches, the formation of carbide buffer layers between diamond and Cu has been most commonly applied to the actual fabrication of Cu/diamond composite. Li *et al.* obtained the composite with an enhanced thermal conductivity of 716 W/m-K by coating Ti on diamond particles (diamond size 70 μm, 65 vol%)[7]. Shen *et al.* fabricated the composite with an enhanced thermal conductivity of 726 W/m-K by forming $MoC_2$ layer (diamond size 100 μm, 65 vol%)[8]. Abyzov *et al.* enhanced the thermal conductivity of the composite by forming CrC or WC and obtained thermal conductivity of 770 W/m-K (diamond size 180 μm, 61 vol%)[9]. Formation of other kinds of carbide buffer layers like $B_4C$[10] and ZrC[11] also help to fabricate high thermal conductivity Cu/diamond composite. Note that the values of thermal conductivity introduced above were measured at room temperature.

Although an appropriate carbide buffer layer can improve the thermal conductivity of the Cu/diamond composite, the thickness and component of these buffer layers are hard to control during the high-temperature fabrication process, where the diamond surface reacts with the buffer layer[12]. This would give rise to alloying of the buffer layer and diamond with a thickness corresponding to their inter-diffusion length. Although a thicker alloyed layer improves the binding of diamond and Cu, since the thermal conductivity of the alloyed layer is relatively low, the partially alloyed buffer layer suppresses the effective TBC[13].

More recently, the self-assembled monolayer (SAM) technique[14], which can form a uniform organic layer (thickness 1-2 nm) of alkyl chain molecules, has been applied for TBC enhancement. Many molecular dynamics (MD) simulations studies have verified the function of SAM in improving the TBC. Luo *et al.* indicated that thermal transport across gold (Au)-SAM-Au[15][16] and GaAs-SAM-GaAs[17] could be promoted by bridging the mismatch of the vibrational density of states (vDOS) of the interface by SAM. Hung *et al.* studied the thickness of the SAM layer on TBC determination at Cu/SAM/diamond interface, discovering that the thickness-dependent spectral transmission of low-frequency phonons predominate the TBC[18]. Experimental studies using the TDTR method also clarified the effect of SAM on TBC enhancement. Sun *et al.*[19] enhanced the TBC by six-fold (from 28 MW/m$^2$-K to 169 MW/m$^2$-K) through functionalizing the Au/polymer interface with proper SAM to bridge the vibrational mismatch of the pristine interface. Losego *et al.*[20] and O'Brien *et al.*[21] achieved a doubled (from 36 MW/m$^2$-K to 65 MW/m$^2$-K, Au/ quartz interface) and 4.7 times (from 30 MW/m$^2$-K to 430 MW/m$^2$-K, Cu/SAM/SiO$_2$) enhanced TBC, respectively, by changing the end group of SAM from methyl to thiol, which increases the interfacial adhesion and thus the phonon transmission.

   The progress in the SAM functionalization to enhance TBC, as well as the stability of SAM structure at high temperature[22] motivate applying well-defined SAM to the fabrication of high thermal conductivity Cu/diamond composite. However, despite the progress made in understanding the influence of end group and chain length of SAM on TBC, there is a lack of experimental studies of the effect of SAM morphology (i.e., coverage, thickness, and chain-molecules ordering of SAM), which is sensitive to the formation process especially in scalable production that is essential for industrial application. Therefore, issues concerning the SAM morphology in fabricating high thermal conductivity composite materials still remain to be uncovered.

Here, a systematic study combining the well-controlled model experiment, MD simulation, and composite fabrication/characterization is performed, aiming at developing high thermal conductivity Cu/diamond composite with the SAM functionalization at the interface (Fig. 1 (a)). To understand the effect of SAM to promote the thermal conductivity, we adopt a model system of a Cu deposition layer on a SAM modified diamond substrate to provide detailed characterizations of SAM morphology and the resulting TBC. By performing SAM functionalization on the diamond substrate (for the model system) and particles (for the composite fabrication) concurrently in the same reaction container, morphologies of SAM and corresponding TBC are ensured to be comparable for both cases. Several SAM formation methods are adopted to form SAM with different morphologies. By verifying the correlation between the detailed SAM morphology and the resulting TBC with the assistance of MD simulations, we elucidate the heat conduction mechanism for different SAM morphologies. Consequently, the formation of high-quality SAM gives rise to considerable enhancement of TBC in the model experiment. Applying the best SAM formation condition realizing the highest TBC onto the diamond particles facilitates the fabrication of actual composite with pronounced improvement in thermal conductivity.

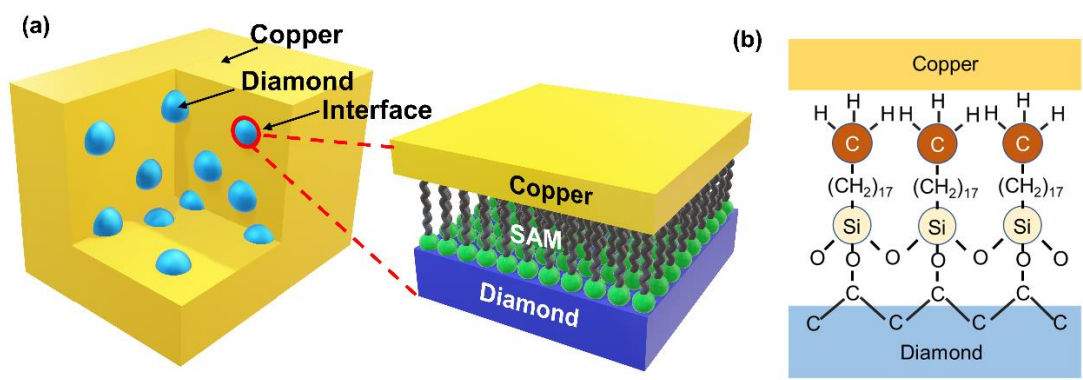

Figure 1. Schematics of (a) copper/SAM/diamond composite and the zoomed up morphology of the SAM-functionalized interface; (b) Chemical structure of copper/SAM/diamond interface.

2. METHOD

**2.1 Formation of SAM**

Samples #1~#6 were prepared with parameters shown in Table S1. The formation of SAM was carried out by two steps. For the 1$^{st}$ step, the diamond sample was immersed into the piranha solution (30% $H_2O_2$ + 70% $H_2SO_4$) for 2 hours following by ultrasonicate washing using deionized water, acetone solution in sequence to obtain hydroxyl-end surface. For the 2$^{nd}$ step, based on our separate experimental study on the effect of end group and alkyl-chain length on TBC of Cu/diamond interface, which will be presented elsewhere soon, suggesting the superiority of octadecyltrimethoxysilane (chain length ~21Å, -$CH_3$ end group) in enhancing TBC, octadecyltrimethoxysilane was used for SAM formation in the present study (Fig. 1 (b)). Both gas-phase and liquid-phase process was used for SAM formation. For the gas-phase method, 30 µL silane, and diamond sample (1 g diamond particles/diamond substrate) were sealed inside a nitrogen-filled container and then heated to 150 °C for 2 hours. For the liquid-phase method, the diamond sample was first immersed in a 30 mL toluene solution (99.5%, Wako) with 30 mL silane and 45 mL triethylamine for 2 or 48 hours in a sealed container. Afterward, The SAM-functionalized diamond sample prepared by both methods were washed by toluene, acetone solution using ultrasonicate in sequence in case of eliminating the physically adsorbed silane.

**2.2 Morphology of SAM**

To characterize the morphology of SAM, deionized water contact angle measurement, ellipsometry measurements, X-ray photon spectroscopy (XPS) was carried out. The thickness and contact angle of SAM was evaluated by using the substrate sample prepared in the same manner and the same reaction vessel as diamond particles. The contact angle, which is decided by the surface energy of the substrate, strongly depends on the morphology of the surface. Therefore, it is commonly used to evaluate SAM morphology. MARY-102 (Five lab) ellipsometer, which is equipped with a 633 nm laser, and a fixed incident angle of 70 degrees, was used to measure the SAM thickness. 25 spots were measured for individual samples to obtain the thickness and confirm the uniformity of SAM layer. The SAM thickness was obtained by fitting the measured psi and delta to the theoretical curve calculated with the air/SAM/substrate (diamond or silicon) structure (Fig. S1). The refractive index and extinction coefficient of each layer was summarized in Table S3. Herein, the thickness measurement was also performed for SAM that concurrently formed on silicon substrates to ensure the validity of the result, since the psi varies slightly with the thickness of SAM on diamond (Fig. S1 (b)), but varies significantly for that on the silicon substrate (Fig. S1 (a)).

The XPS measurement (Ulvac, PHI-5000 Versaprobe iii) was used to evaluate the coverage of SAM on the surface of diamond particles. Here, we focused on the XPS spectrum around the carbon 1s (C-1s) and silicon 2p (Si-2p) peaks. Since C-1s peaks mostly originate from diamond particles, with Si-2p peaks originating from the alkyl chain molecules, it is possible to evaluate the SAM coverage by the intensity ratio of Si-2p to C-1s peaks ($I_{Si}/I_C$). By comparing the $I_{Si}/I_C$ of 6 different samples, we can obtain the relative coverage of SAM.

## 2.3 TDTR measurement

TDTR was used to measure the TBC of the SAM modified Cu/diamond interface. TDTR is a pump-probe measurement method to solve the nanoscale heat conduction issue[23]. In this measurement, pulse laser (frequency~80 MHz, pulse width~140 fs) with a wavelength of 800 nm is split into two ways, with one way followed by a wavelength conversion to 400 nm for pumping and the other way for probing. The pulse laser signal is modulated with specific frequencies (11.05 MHz in the present study) to avoid the noise as well as to control the penetration depth. Whenever the aluminum (Al) transducer layer on top of the sample absorbs the energy of the pump laser, there is a temperature decay on the top of the transducer that can be probed by the probe laser. By varying the interval time between the pump and probe laser using a light path delay stage for probe laser, we can obtain the continuous temperature decay profile and hence the targeted thermal transport parameters. Here, we used the sample with Al/Cu/diamond structure to measure the TBC of Cu/diamond interface ($G_{Cu-DI}$). The Al (~60 nm) and Cu (~20 nm) was deposited by EB vacuum evaporation, where the Al layer acts as the transducer for the thermal reflectance signal. For this structure, because the TBC of Al/Cu interface ($G_{Al-Cu}$) is extremely small, making the corresponding sensitivity ignorable (Fig. S2 (a)), the Al/Cu layers can be regarded as single layer[24]. This can simplify the heat transfer model to a two-layer structure (Table S2 summarizes the parameters of the structure). For this structure, the sensitivity calculation was also carried out (Fig. S2 (b)), showing a high sensitivity for the TBC of Cu/diamond interface ($G_{Cu-DI}$), thus the high reliability of the TBC value.

## 2.4 MD simulation

Non-equilibrium molecular dynamics (NEMD)[25] simulations were carried out to study the TBC across Cu/SAM/diamond surfaces. The generation of Cu and diamond surfaces follows the procedure of our previous study[18] to form a 5.0×5.0×30.0 nm$^3$ slab for each surface. The diamond surface was fully hydroxylated initially. The hydroxyl groups were then used as the starting points to graft the alkylsilane(-OSi(OH)$_2$(CH$_3$)$_{17}$CH$_3$). Seven coverage, 0.72, 0.96, 1.40, 2.24, 3.12, and 3.92 molecules/nm$^2$, were examined to understand the effect of SAM coverage on interfacial heat transfer.

The interactions of Cu atoms were described by using the embedded atom method (EAM) potential.[26] The interactions of carbon atoms within the diamond surface were described by using the Tersoff potential.[27] The interactions between Cu and diamond surfaces were adopted from the previous study.[28] The force field parameters of hydroxyl groups and alkyl chains on the diamond surface were obtained from Summers *et al.*[29], which is based on the OPLS all-atom model[30]. The parameters of the force field were summarized in Table S4-S7. The geometric combination rule was applied for the parameters of van der Waals interactions between different atoms. The Coulomb interaction was treated using the particle-particle-particle mesh (PPPM) Ewald summation method[31] for slab geometries.[32] The equations of motion were integrated with a time step of 0.5 fs. All the simulations were performed using LAMMPS[33] molecular dynamics package and visualized with the PyMOL[34] software.

The system was initially relaxed in the canonical ensemble (NVT) at 300 K using Nose-Hoover[35,36] thermostat to ensure the system temperature. After that, two 0.5 nm thick layers at the two ends in the *z*-direction were fixed to stabilize the system. The system size in the z-direction was adjusted gradually according to the normal stress to remove the normal stress imposed at the interface. After the systems reached equilibrium, the NEMD simulations with constant heat exchange algorithm[37] were performed in the microcanonical (NVE) ensemble. The heat transfer was established by adding specified amounts of energy to the heat source region and subtracting the same amounts of energy from the heat sink region. The heat source and heat sink regions were defined as a 1.0 nm thick layer next to the fixed layers. The temperature difference at the interface, $\Delta T$, can be obtained from the temperature profile of the intermediate region. The value of TBC, $G$, can be evaluated by $G=J/\Delta T$, where $J$ is the magnitude of heat flux, which was 0.6 GW/m$^2$. A 5 ns production run was performed to obtain a steady temperature profile. A linear regression of the temperature profile of each surface was applied to determine the value of $\Delta T$.

## 2.5 Composite fabrication

Cu/diamond composite with a 50% diamond volume fraction was fabricated by plasma sintering (ELENIX, ED-PAS). The particle size of Cu particles (Sigma-Aldrich) is around 45 μm; diamond particle size is around 210 μm with thermal conductivity of 1900 W/m-K (Henan Famous Industrial Diamond Co. Ltd., China). Sintering starts from a short pulse plasma treatment (150 seconds) in order to eliminate the oxide layer on Cu particles, followed by the primary sintering process using direct current. The sintering pressure was fixed to be 70 MPa, with the vacuum pressure of the chamber below 5 Pa. Thermal conductivity of the composite was measured by laser flash (NETZSCH, LFA447), with density measured based on Archimedes' principle, heat capacity calculated from the theoretical value of Cu and diamond.

## 3. RESULT AND DISCUSSION

Here, we briefly describe the correlation of the coverage, thickness, and chain-molecular ordering of SAM, which becomes important in the following discussion. In the case of close-packed alkyl chain molecules (high SAM coverage), as illustrated in Fig.1 (b), owing to the strong interchain repulsive force, the alkyl chain molecules are straight and stand perpendicular to the diamond surface spontaneously, forming an ultra-thin SAM layer, whose thickness approximates to the length of a single octadecyltrimethoxysilane chain molecule. In contrast, when the alkyl chain molecules are loose-packed (low SAM coverage), the morphology of SAM becomes relatively complicated. Because of the larger inter-chain distance, the repulsive force interaction becomes weaker, making the chain molecules bent and tangled. Although there is no experimental study investigating the influence of this poor ordering and the resultant small thickness of the SAM layer[38] on TBC, some simulation studies[39] have pointed out possible negative consequences on the heat transfer through the SAM bridges.

Therefore, for the composite of Cu and SAM-functionalized diamond particles, whose interface area is extensive and hard to control, it is a critical issue to elucidate the correlation between TBC and SAM morphology experimentally. To evaluate the morphology of SAM, we conducted detailed characterizations of the morphology indicators like the coverage, thickness, and ordering. To make the TBC measured on the model system comparable to that in the composite, we performed SAM functionalization for diamond substrate (in the model system used for TDTR, contact angle, and ellipsometer measurements) and diamond particles (in the actual composite used for XPS and Fourier transform infrared spectroscopy (FT-IR) measurements) at the same time. To actively control the morphology of SAM, we performed SAM formation by two commonly used gas-phase[40] and liquid-phase[14] methods under different detailed experimental conditions (sample #1~#6). The SAM coverage, which is defined as the number of alkyl chain molecules per surface area on diamond, was conducted by XPS measurement. Figure 2 (a) shows the SAM coverage indicated by the intensity ratio of Si-2p to C-1s peaks ($I_{Si}/I_C$). By comparing the $I_{Si}/I_C$ of 6 different samples (Fig. 2 (b)), we found that in the case of sample #6 prepared by the combination of piranha-solution pretreatment and liquid-phase method with long process time (48h), $I_{Si}/I_C$ is significantly enhanced by four times compared with that of sample #2 prepared by the gas-phase method with short process time. The FT-IR was used to characterize the SAM on sample #6, where sharp peaks representing the C-H vibration mode (2850 cm$^{-1}$ and 2920 cm$^{-1}$) (Fig. 2 (c)) was observed, also indicating the existence of SAM on the diamond surface.

The water contact angle and ellipsometer measurements were conducted for the representative diamond-substrate samples #1, #4, and #6 to confirm the coverage and to evaluate the ordering of SAM (Fig. 2 (d)). Compared with the small contact angle around 45° in the case of the pristine sample (sample #1), the contact angle increases to 60° in the low SAM coverage case (sample #4), and further exceeds 100° in the high SAM coverage case (sample #6). Since the methyl-ends functional groups of octadecyltrimethoxysilane are hydrophobic, the increasing contact angle indicates the increase of the coverage and the improvement of the ordering[41]. This result agrees well with the coverage measured by XPS. However, since the alkyl chain is also hydrophobic, extra physically adsorbed alkyl chain molecules may also lead to an increase of the contact angle.

To further clarify the morphology of SAM, eliminating the possibility of the physical adsorption, the thickness of the SAM was measured using the ellipsometer (Fig. 2 (d)). Sample #6 with high SAM coverage shows a thickness of around 19 Å, similar to that of the close-packed model described above, whose thickness is around 21 Å. Combined with the measured large contact angle around 100°, SAM in sample #6 was verified to be monolayer with good molecular ordering. On the other hand, the measurement of the low SAM overage (sample #4) case gives a smaller thickness around 15 Å and, together with the smaller contact angle, reveals the relatively poor ordering of SAM. These results rule out the possibility of SAM physical adsorption and identify the difference between SAM morphologies in the samples, which enables us to discuss their effect on TBC.

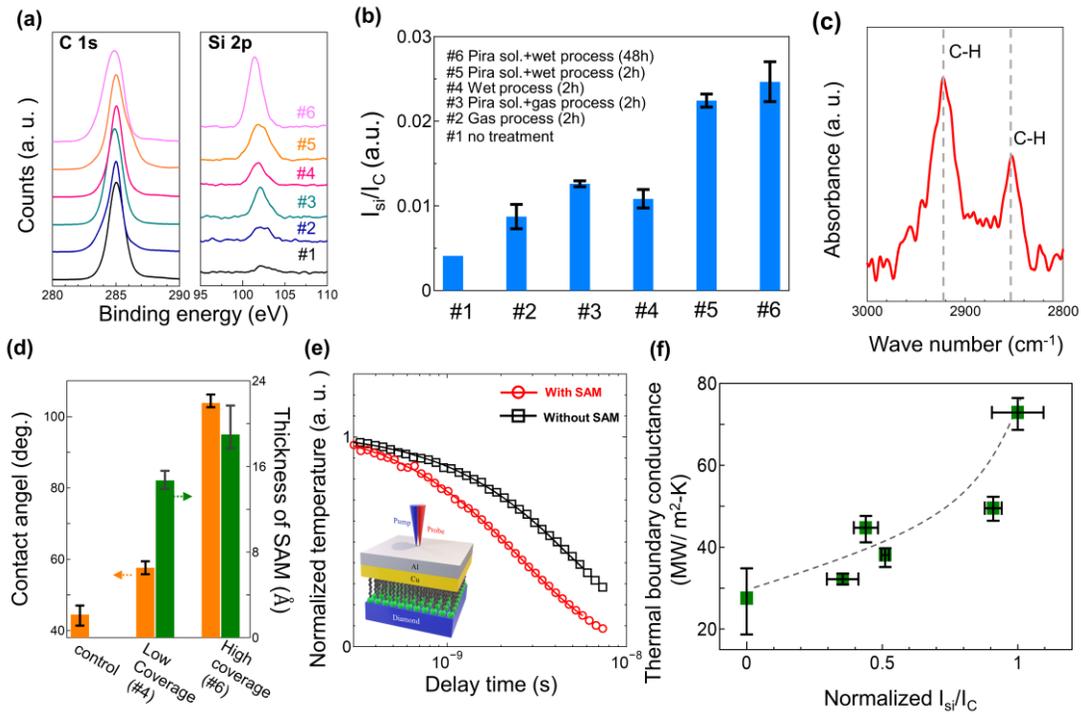

Figure 2. Results of the model experiments. (a) XPS spectrum of C 1s and Si 2p and (b) intensity ratio of Si and C1s peaks ($I_{Si}/I_C$) for samples by different SAM functionalization methods (sample #1~#6); (c) FT-IR spectrum of sample #6; (d) Contact angle and SAM thickness of samples #1, #4 and #6; (e) Temperature decay profiles for the samples with (sample #6) and without SAM (sample #1) measured by TDTR, markers (squares and circles) are experimental data, and curves correspond to the fitting results to the physical (heat conduction) model; (f) Thermal boundary conductance of samples with different SAM coverages (Dashed curve is the guide for the eye).

The TDTR measurement was carried out for samples #1~#6. The inset figure in Fig. 2 (e) shows the schematic of the model experimental system. Figure 2 (e) shows the temperature decay profiles for samples with (sample #6) and without (sample #1) SAM functionalization (The temperature decay profiles for samples #2~#5 is shown in Fig. S3). There is a distinct difference in the temporal profiles where the decay is considerably faster in the SAM-functionalized case, indicating that bridging the Cu/diamond interface with SAM can significantly enhance TBC. Moreover, by plotting the TDTR-measured TBC for sample #1~#6 as a function of relative coverage represented by $I_{Si}/I_C$ (Fig. 2 (f)), we found that TBC can be enhanced by increasing the SAM coverage. Besides, the trend is nonlinear, with the slope of TBC increasing as the SAM coverage increases. As the previous study[18] has demonstrated, the SAM can bridge the phonon vibration spectra of the two materials with distinctly different vibrational properties[42], like Cu and diamond, it is possible to attribute the TBC enhancement to the increasing number of SAM bridges.

One issue here is whether to view this bridging effect as one SAM molecule giving rise to one phonon transport channel as in the often discussed 'parallel-channel model'[16]. In this model, each SAM molecule is assumed to act as an isolated channel due to the weak van der Waals interchain interactions, ignoring the effects of interchain thermal energy transport, interchain phonon scattering, and collective interchain vibrational modes. However, this would result in a linear increase of TBC with increasing SAM coverage and contradicts with our experimental result. This gap between the parallel-channel model and the present experimental result may result from the difference in the coverage ranges. The previous theoretical studies mainly focused on the relatively high coverage cases, where molecular ordering did not explicitly change with coverage, while the present experimental work studied a broader range of coverage, including the low coverage cases. In such an extended range of coverage, the variation of molecular ordering becomes much more significant. This assumption was supported by the MD study on the morphology variation with coverage[38], which suggested a significant molecular ordering variation in the case of low coverage, whereas slight change under high coverage. Besides, according to the studies of heat transfer through one-dimensional systems like single polymer nanofiber[43] and polyethylene single-molecule[39], the conformation of the chain molecules was demonstrated to critically influence the intrachain phonon transport and the vibration modes. These factors may be the cause of the experimentally observed nonlinear dependence of TBC on the coverage.

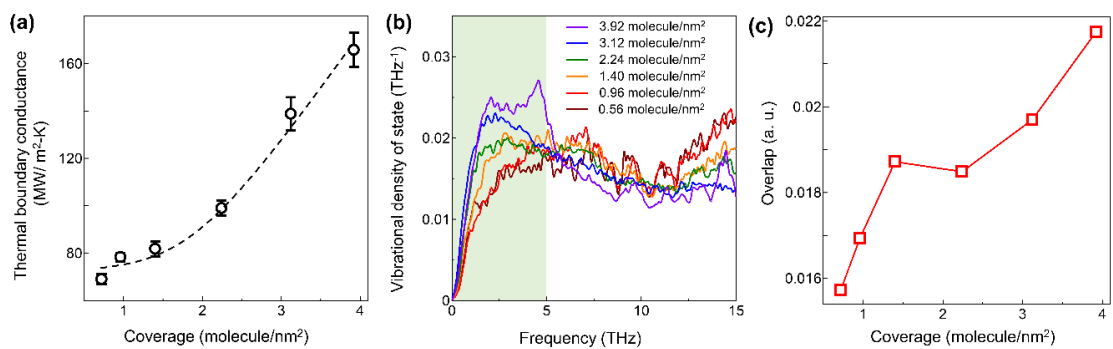

Figure 3. Molecular dynamics simulation results. (a) Thermal boundary conductance (the dashed curve is the guide for the eye), (b) vibrational density of states (0-15 THz) of SAM, and (c) corresponding vibrational spectrum overlap of copper and SAM with different coverages.

To further explore the cause of the coverage dependence of TBC, we carried out NEMD simulations. The schematics and corresponding temperature profiles of the representative simulation system are shown in Fig. S4. The resultant values of TBC for different SAM coverages were summarized in Fig. 3(a). Compared with the interface without SAM, the enhancement in TBC, from 24.8 to 68.9 MW/m$^2$-K, can be achieved even at the lowest SAM coverage, 0.72 molecule/nm$^2$. With further increase of the SAM coverage, TBC keeps increasing at an elevating increase rate (Fig. 3(a)). The MD simulations capture the experimentally observed nonlinear increasing trend with the coverage for SAM-functionalized cases.

Using the above MD simulations, we evaluated the phonon transport in terms of matching of the vibrational spectra, which determines the effect of SAM bridging. For this purpose, the vDOS, computed by taking the Fourier transform of normalized velocity autocorrelation functions of atoms,[44] were calculated for different SAM coverages. Figure S5 shows the calculated vDOS for different components with different SAM coverages. For the case with only hydroxyl groups (0 coverage), the large vDOS mismatch between the diamond and Cu results in the low TBC. With the addition of SAM, whose vDOS covers that of both Cu and diamond, the effect of bridging over the vibrational mismatch can be enhanced. Herein, since the temperature jump at the Cu/SAM interface is much larger than that inside the SAM and at the diamond/SAM interface (Fig. S4 (b)), the Cu/SAM interface predominately determines the TBC. Besides, we observed that the vDOS of SAM shifts to the low-frequency regime (0−5 THz) with the increasing SAM coverage (Fig. 3(b)). The increase of low-frequency phonon modes inside SAM enhances the vibrational matching between SAM and Cu, whose vDOS mainly distributes within 0-8 THz. This improvement of spectral overlap between Cu and SAM[45] with increasing coverage (Fig. 3(c)) causes the enhancement of TBC. Note here, as the vDOS of SAM were calculated from the normalized velocity autocorrelation function, which means that the vDOS of SAM is not a function of the number of SAM molecules, the increasing vDOS overlap between Cu and SAM indicates the enhancement of phonon transport along individual SAM molecules. This promoted phonon transport should be responsible for the increasing slope of TBC with SAM coverage.

Herein, we give a possible explanation for the variation of vDOS in terms of the SAM morphology under different coverages since the vibration of alkyl chain molecules is sensitive to the interaction with surrounding molecules[46]. We calculated the root-mean-square displacement (RMSD) of all SAM molecules to obtain the average mobility, which was found enhanced with the increase of coverage (Fig. S6). This result matches well with the morphology variation of SAM. Under low coverage conditions, the movement of chain molecules is strongly restricted as the backbones of some alkyl chain molecules attach on the surface of diamond or Cu (insertion (a) in Fig. S6). A model proposed by F. Sun *et al.*[19] in estimating the vDOS can be implemented to describe the variation in such cases. In this model, considering the whole SAM layer as a harmonic oscillator connecting between the Cu and diamond, the vibrational frequency of the SAM was indicated roughly proportional to $\sqrt{F}$, where $F$ is the force constant taking the SAM layer as a whole[19]. Because the force constant of the whole SAM layer is partially decided by the nonbonded (like the van der Waals force) interaction, the direct interaction between the backbone of SAM and the diamond or Cu, whose Young's modulus is typically more than two orders of magnitude higher than that of SAM (~1 GPa)[47], may increase the force constant of SAM. This would lead to the upshift of the vDOS and hence cause the decrease of the population of low-frequency modes, agrees with that in the previous study on the water/SAM/gold system[48]. Following this, in the case of high coverage, the small force constant induced by the interchain interaction between individual SAM molecules[47] (insertion (b) in Fig. S5) may, in turn, induce the downshift of vDOS spectra and hence high population of the low-frequency vibration modes. It should be noted that proof of the explanation for the downshift of vDOS under higher SAM coverage requires more exploration and clarification considering the complexity of interaction in the SAM system and remains open for future studies. Based on the model-experiment and MD simulation studies, the high SAM coverage diamond particles (sample #6) were chosen for the fabrication of high thermal conductivity

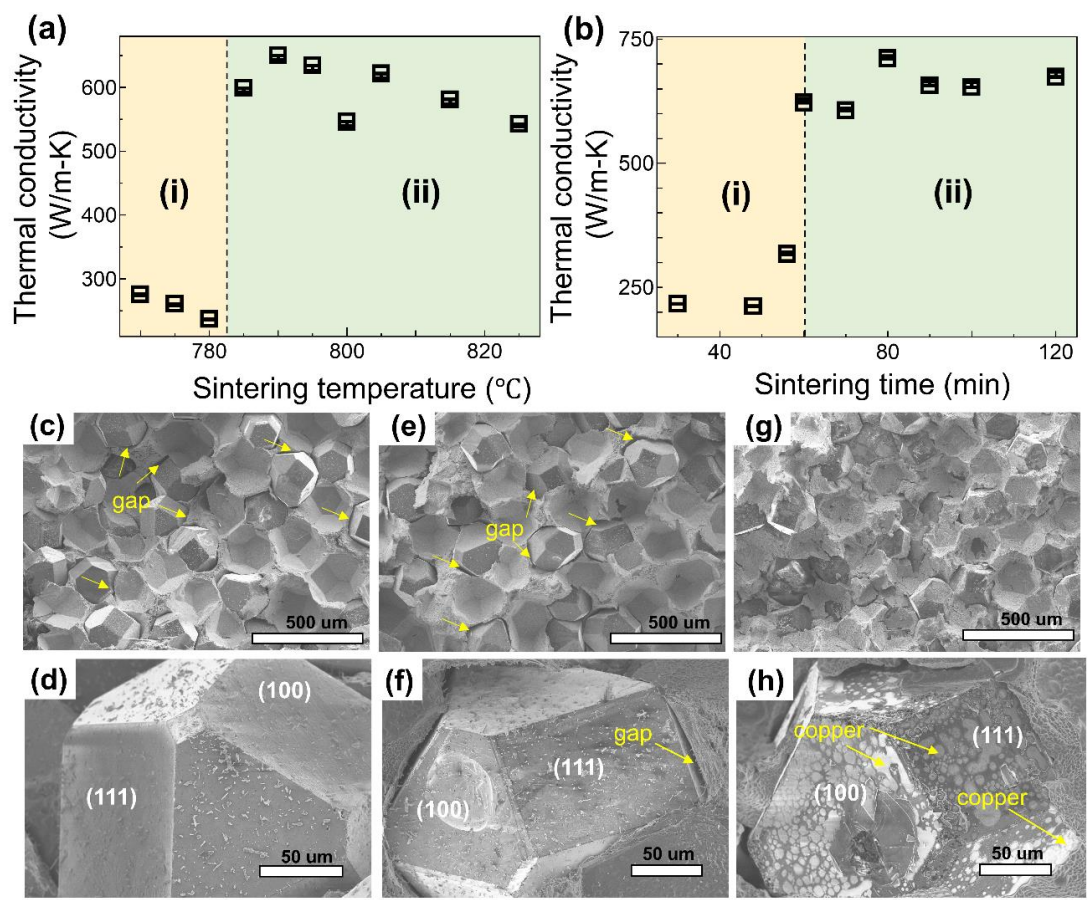

Figure 4. Thermal conductivity of copper/SAM-functionalized-diamond composite with different (a) sintering temperature and (b) sintering time, SEM image of composite sintered under (c)(d) low temperature (770℃, 60 mins), (e)(f) short sintering time (810 ℃, 15 mins), and (g)(h) high temperature, long sintering time (795℃, 60 mins). The arrows on the SEM images are to indicate the gaps between copper and diamond. The arrows in (h) denote some locations of copper on diamond particles.

Cu/diamond composite by the plasma sintering method. Different sintering temperatures and time were used to fabricate the composite samples. Figure 4 (a) shows the dependence of thermal conductivity on the sintering temperature that can be divided into two regions. With increasing sintering temperature from 770 °C, the thermal conductivity is low and remained almost unchanged below 785 °C [region (i)]. However, the thermal conductivity is discontinuously elevated at 785 °C, above which the value is steadily high [region (ii)]. The effect of sintering time is shown in Fig. 4 (b). Similar to the trend of temperature dependence, the thermal conductivity is low for short sintering time [region (i)], and it steeply increases when sintering time is 60 minutes and then saturates quickly. With high-temperature and long-time sintering, thermal conductivity as high as 711 W/m-K was achieved.

As a comparison, the composite was fabricated with pristine diamond particles treated only by piranha solution (pristine-diamond). Unlike that of SAM-functionalized-diamond, for pristine-diamond, with the increase of sintering temperature, the thermal conductivity increases continuously at low temperatures (<800 °C) and saturates at around 800 °C (Fig. S7 (a)). This difference in sintering-parameter dependences can be explained by the difference in the diamond surface chemical state. For pristine-diamond, whose surface is terminated by hydroxy, is highly hydrophilic. Hence, the affinity between intenerated Cu and the pristine-diamond surface is much higher than that of SAM-functionalized-diamond, which has a hydrophobic surface that originates from the methyl end group. Due to the difference in affinity, which is a function of temperature[49], the formation of a tightly adhered interface using SAM-functionalized-diamond requires higher temperature and longer reaction time.

This explanation can be further confirmed by the tensile fracture surface imagine of the composite measured by scanning electron microscope (SEM). For samples fabricated under low temperature and short sintering time [region (i) in Fig. 4 (a)(b), respectively] using SAM-functionalized-diamond, many large gaps between diamond and Cu matrix were observed because of the weak binding (Fig. 4 (c)-(f)). When the sintering temperature and time were increased to above 800 °C and 1 hour [region (ii) in Fig. 4 (a)(b)], the binding between Cu and diamond is well promoted and result in fewer gaps generated by the tensile fracture. Meanwhile, we observed some Cu on the diamond surface originated from the transcrystalline fracture of the Cu matrix, which only happens when the interfacial bonding is more robust than the fracture stress of Cu matrix (Fig. 4 (g)-(h)).

Furthermore, despite the reported high thermal stability[22], since the difference in the detailed heating parameters like heating time and vacuum degree may influence the thermal stability of SAM[50]. It is necessary to confirm the existence of SAM after the high-temperature process used in this study. The XPS (Fig. S8 (a)(b)) results show no perceptible difference in the Si-2p peaks for SAM-functionalized-diamond particles before and after the high-temperature process (800 °C, 1 hour), with a negligible difference observed in the $I_{Si}/I_C$ (Fig. S8 (c)). This result indicates the survival of SAM after the high-temperature sintering process.

To estimate the TBC of the composite, further analysis was performed using the effective medium theory (EMT)[51], which was proven to be effective in modeling the thermal conduction in alkyl chain cross-linking nanocrystal composite materials[52][53]. Many theoretical models have been developed to describe the impact of the TBC on the thermal conductivity of composites[54]. Among those, the model proposed by Nan. *et al*.[1] is usually used for the composite with high thermal conductivity and shows good accordance with the experiment[55]. Moreover, to account for the impact of pores on the effective thermal conductivity of the matrix, we combine this model with a two-step approach proposed by Chu. *et al*.[56](Details of the model and calculation are shown in Supporting Information). The TBC of diamond composite fabricated by SAM-functionalized-diamond and pristine-diamond were calculated respectively. For the SAM-functionalized-diamond case, the TBC is 82 MW/m$^2$-K, while that of the pristine-diamond case is 30.5 MW/m$^2$-K. These results are slightly higher than that measured by TDTR, whose value is 27.5 MW/m$^2$-K and 72.9 MW/m$^2$-K, respectively, but still remain to be in reasonable agreement (Fig. 5(a)), indicating that the approach to scaling the nanointerface in Cu/SAM/diamond model system to the composite using plasma sintering is effective.

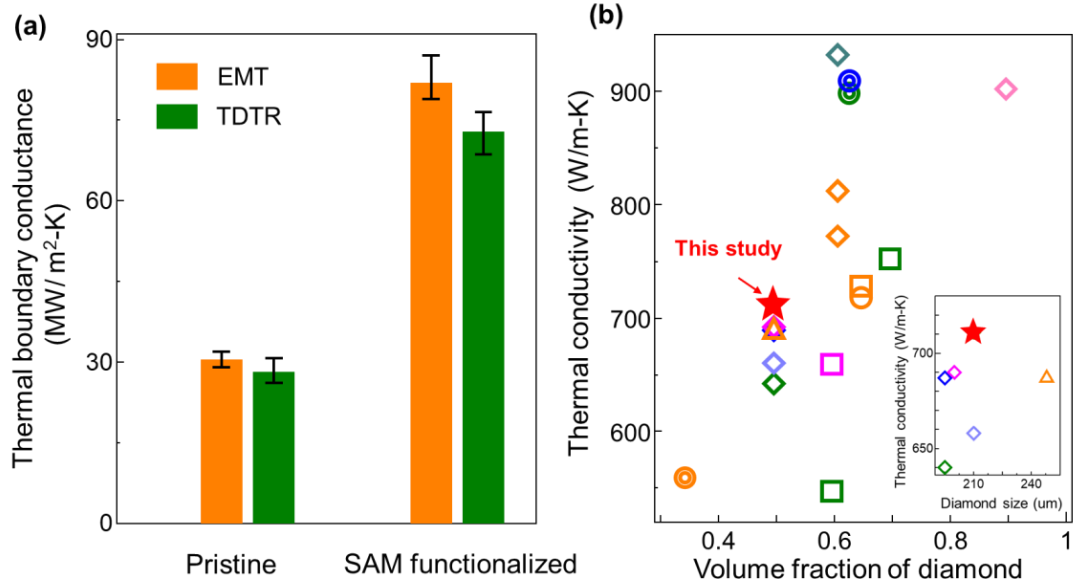

Figure 5. (a) Thermal boundary conductance of pristine- and SAM-functionalized copper/diamond interface measured by time-domain thermoreflectance (TDTR) and calculated by effective medium theory (EMT) using thermal conductivity measured by laser flash, respectively. (b) Summary of the high thermal conductivity copper/diamond composite with different diameter and volume fraction of diamond in studies thus far (different marks represent diamonds with different sizes, ○: ~100 μm, □: 100~150 μm, ◇: 150~230 μm, △:230~250 μm ◎: 400 μm ~)

By applying SAM functionalization on Cu/diamond composite fabrication, we succeeded in enhancing the thermal conductivity to as high as 711 W/m-K. Here, we compared our results with the previous studies (Fig. 5(b))[7,8,60–65,9–13,57–59]. Since the thermal conductivity is particularly sensitive to the diamond size and volume fraction, which are critical for deciding the material cost, although the highest experimentally achieved thermal conductivity is around 900 W/m-K[11], it is only meaningful to compare the performance of the composite fabricated with similar diamond volume fraction and size within a similar range. Here, we extracted the data from studies with the same volume fraction (50 vol%) and similar diamond sizes (200~250 µm), as summarized in the inset of Fig. 5 (b). This illustrates that the thermal conductivity of the composite fabricated in this study is higher than the previous studies, revealing the advantage of SAM functionalization on composite fabrication.

## 4. CONCLUSION

We developed high thermal conductivity Cu/diamond composite using SAM functionalization by systematically performing the model experiment and MD simulation. The influence of SAM morphology on determining TBC of Cu/diamond interface was identified by detailed morphology characterization and TDTR measurement on the model experimental system. We found a nonlinear increase of TBC with increasing SAM coverage, where the slope is growing for high coverage. Assisted by MD simulation, we conclude that it is because the effect of SAM to bridge the vibrational mismatch varies with the SAM morphology. Contributed by the increase of the number of 'phonon channels' and the phonon transmission via individual 'channels' with the increase of SAM coverage, the increasing trend of TBC is elevated. This controlled functionalization of Cu/diamond nanointerface can give TBC as high as 73 MW/m$^2$-K. By scaling this nanointerface via a well-controlled plasma sintering process, we obtained a Cu/diamond composite with high thermal conductivity of 711 W/m-K. This study not only provides a scientific perspective on the mechanism of heat conduction across SAM-functionalized nanointerface but also reveals the advantages of the SAM functionalization for the application of high thermal-conductivity composite materials.

**Author Contributions**

J. S. conceived the idea and supervised the whole project. B. X. fabricated the sample. B. X. and R. G. performed the experimental measurements. T. K. and J. C. support with the silane coupling and characterization. S-W. H and S. H. performed the simulations with support from C. S and F-R. C.. B. X., S-W. H., and J. S. wrote the manuscript. All authors have participated in the discussion.

**Acknowledgements**

Some of the equipments used in this study were supported by JSPS KAKENHI (19H00744) and JST CREST(JPMJCR16Q5, JPMJCR19I2).

# Scalable monolayer-functionalized nanointerface for thermal conductivity enhancement in copper/diamond composite

*Bin Xu[1], Shih-Wei Hung[2], Shiqian Hu[1], Cheng Shao[1], Rulei Guo[1], Junho Choi[1], Takashi Kodama[1], Fu-Rong Chen[2], Junichiro Shiomi[1*]*

[1]Department of Mechanical Engineering, The University of Tokyo, Tokyo, Japan, E-mail:

[2]Department of Materials Science and Engineering, City University of Hong Kong, Hong Kong, China

*Email: shiomi@photon.t.u-tokyo.ac.jp

## Section 1. Force field of MD simulation

The non-bonded parameters in the work of Guo *et al.* [1] were used to describe the non-bonded interactions of copper and diamond surfaces. The parameters implemented in the work of Summers et al.[2] for the alkylsilane chains were adopted. The potential functions were given by

$$U_{\text{total}} = \sum_{\text{nonbonded},i} \sum_{i<j} \left\{ 4\varepsilon_{ij} \left[ \left(\frac{\sigma_{ij}}{r_{ij}}\right)^{12} - \left(\frac{\sigma_{ij}}{r_{ij}}\right)^{6} \right] + \frac{q_i q_j}{4\pi\varepsilon_0 r_{ij}} \right\} + \sum_{\text{bonds}} k_r (r - r_0)^2 + \sum_{\text{angles}} k_\theta (\theta - \theta_0)^2 + \sum_{\text{dihedrals}} \{V_1[1 + \cos(\varphi)] + V_2[1 - \cos(2\varphi)] + V_3[1 + \cos(3\varphi)]\} \quad (1)$$

The total potential functions consist of the Lennard−Jones and Coulomb terms for intramolecular and intermolecular non-bonded interactions and bond stretching, bond angle bending, and dihedral angle torsion terms for bonded interactions. The parameters of each interactions were listed in the following tables.

## Section 2. Effective medium theory

The effective medium theory (EMT) proposed by Nan. *et al.*[3] is used to estimate the effective thermal conductivity of the diamond/copper composite ($K$):

$$K = K_m \frac{K_p(1+2\alpha) + 2K_m + 2f[K_p(1-\alpha) - K_m]}{K_p(1+2\alpha) + 2K_m - f[K_p(1-\alpha) - K_m]} \quad (2)$$

$$\alpha = K_m / (G \cdot r) \quad (3)$$

where $K$ is the thermal conductivity; subscripts *eff*, *m*, and *p* denotes the composites, matrix, and filler, respectively; $f$ is the volume fraction of the filler, $r$ is the radius of diamond filler, and $G$ is the thermal boundary conductance (TBC) of diamond/copper interface.

Moreover, to account for the effect of pores in the copper matrix, we used a two-step approach proposed by Chu *et al.*[2], which enables calculating the thermal conductivity of composite with three components. Since all the pores exist inside the copper matrix, the effective thermal conductivity of the matrix can be obtained considering a composite material comprising pores and copper. Then, the resultant effective thermal conductivity of the matrix was used to calculate the thermal conductivity of copper/diamond composite with pores by EMT (Fig. S9 (a)).

Due to the low thermal conductivity of air ($\kappa$~0.026 W/m-K) inside the pores, the TBC between the pores and the copper is negligible. This also makes the impact of the size of the pore insignificant to the thermal conductivity. Therefore, a Hasselman-Johnson model can be used to predict the effective thermal conductivity of the copper matrix ($\kappa_m^{eff}$):

$$\kappa_m^{eff} = \kappa_m \frac{1-f'_{pore}}{1+0.5 f'_{pore}} \quad (4)$$

The effective porosity is expressed as $f'_{pore} = f_{pore}/(1-f)$, where the $f_{pore}$ and $f$ are the porosity and the volume fraction of the diamond filler, respectively. Figure S9 (b) indicates the effective thermal conductivity of the copper matrix with a porosity of 0~8%, which is the case in this study. With the increase of the porosity, the effective thermal conductivity of copper decreases and eventually reaches 311.1 W/m-K under porosity of 8%, reduced by 22.2% from the bulk value. We then replaced the $K_m$ with the $\kappa_m^{eff}$ in Eq. (2) and eventually obtained the thermal conductivity ($K$) of the copper/diamond composite (Fig. S9 (c)):

$$K = \kappa_m^{eff} \frac{K_p(1+2\alpha)+2\kappa_m^{eff}+2f[K_p(1-\alpha)-\kappa_m^{eff}]}{K_p(1+2\alpha)+2\kappa_m^{eff}-f[K_p(1-\alpha)-\kappa_m^{eff}]} \quad (5)$$

**Tables**

Table S1. Parameters for the formation of SAM in sample #1~#6.

| Samples | Step 1. Pretreatment | Step 2. SAM formation |
|---|---|---|
| #1 | None | None |
| #2 | None | Gas process (2h 150 °C) |
| #3 | Piranha solution treatment | Gas process (2h 150 °C) |
| #4 | None | Wet process (2h) |
| #5 | Piranha solution treatment | Wet process (2h) |
| #6 | Piranha solution treatment | Wet process (48h) + DTT |

Table S2. Parameters of TDTR fitting for the two-layer structure.

|  | Thickness (nm) | Thermal conductivity (W/m-K) | Specific heat (J/m3) |
|---|---|---|---|
| Layer 1 (Al/Cu) | 80 | 150 | $2.4 \times 10^6$ |
| Layer 2 (Diamond) | $3 \times 10^5$ | 1500 | $1.8 \times 10^6$ |

Table S3. Parameters for ellipsometry fitting

| Layer | Refractive index | Extinction coefficient |
|---|---|---|
| Air | 1 | 0 |
| SAM | 1.43 | 0.025 |
| Silicon | 3.820 | 0.198 |
| Diamond | 2.4 | 0 |

Table S4. Non-bonded parameters

| Atom Type | Description | $\varepsilon$ (eV) | $\sigma$ (Å) | $q$ (e) |
|---|---|---|---|---|
| Cu | Copper | 0.16700 | 2.314 | 0 |
| C | Carbon in diamond | 0.00456 | 3.851 | 0 |
| C | Carbon in diamond bonded to hydroxyl group | 0.00456 | 3.851 | 0.265 |
| C | Carbon in diamond bonded to alkylsilane | 0.00456 | 3.851 | 0.468 |
| O | Oxygen in hydroxyl group | 0.00737 | 3.120 | -0.683 |
| H | Hydrogen in hydroxyl group | 0 | 0 | 0.418 |
| Si | Silicon in alkylsilane | 0.00434 | 4.000 | 0.745 |
| C | Carbon in methylene group | 0.00286 | 3.500 | -0.120 |
| C | Carbon in methyl group | 0.00286 | 3.500 | -0.180 |
| H | Hydrogen in methylene or methyl groups | 0.00130 | 2.500 | 0.060 |

Table S5. Bond parameters

| Bond type | $k_r$ (eV/Å$^2$) | $r_0$ (Å) |
|---|---|---|
| O-C (Diamond) | 13.876 | 1.410 |
| Si-O | 13.009 | 1.630 |
| O-H | 23.980 | 0.945 |
| Si-C | 8.673 | 1.850 |
| C-H | 14.743 | 1.090 |
| C-C | 11.621 | 1.529 |

Table S6. Angle parameters

| Angle type | $k_\theta$ (eV/rad$^2$) | $\theta_0$ (°) |
|---|---|---|
| Si-O-C (Diamond) | 0.867 | 145.0 |
| O-Si-O | 2.602 | 110.0 |
| O-Si-C | 2.602 | 100.0 |
| Si-O-H | 1.031 | 122.9 |
| Si-C-C | 1.321 | 120.0 |
| H-C-H | 1.431 | 107.8 |
| C-C-H | 1.626 | 110.7 |
| C-C-C | 2.530 | 112.7 |
| H-O-C (Diamond) | 2.385 | 108.5 |

Table S7. Dihedral parameters

| Dihedral type | $V_1$ (eV) | $V_2$ (eV) | $V_3$ (eV) |
|---|---|---|---|
| H-C-C-H | 0 | 0 | 0.00650 |
| C-C-C-H | 0 | 0 | 0.00650 |
| C-C-C-C | 0.02819 | -0.00108 | 0.00434 |

**Figures**

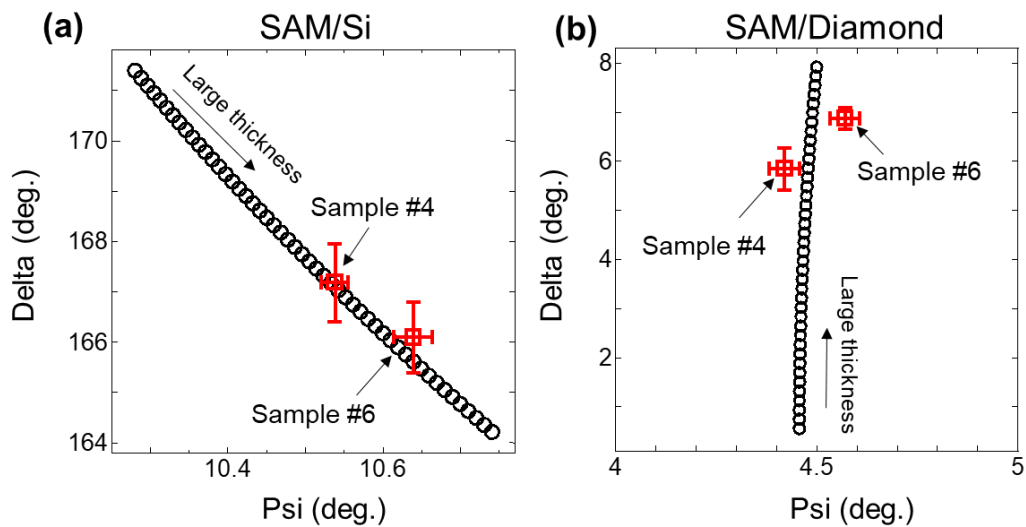

Figure S1. Fitting the thickness of SAM on (a) silicon, and (b) diamond substrate. The red mark is the experimentally measured psi and delta, the black circles are the theoretical curve for SAM with thickness ranges from 0.5 to 25.5 Å; the interval of each circle mark is 0.5 Å.

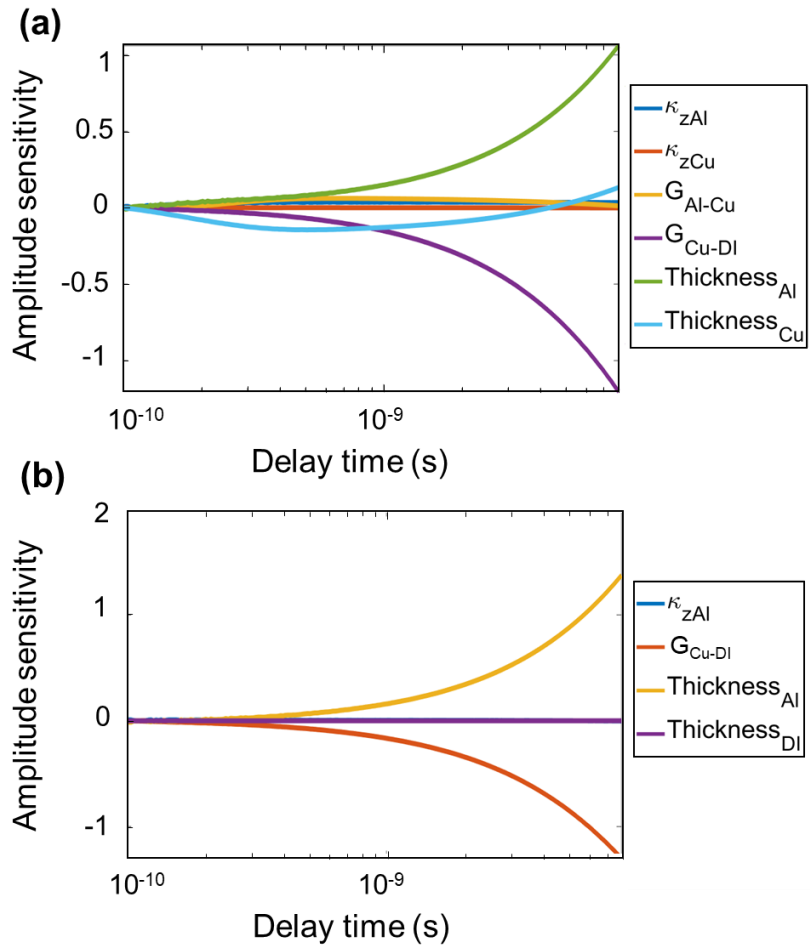

Figure S2. TDTR sensitivity calculation of the (a) aluminum/copper/diamond three-layers structure and (b) (aluminum+copper)/diamond two-layers structure. κ represents the thermal conductivity, G represents the thermal boundary conductance, with subscript Al, Cu, DI correspond to alumina, copper, and diamond, respectively.

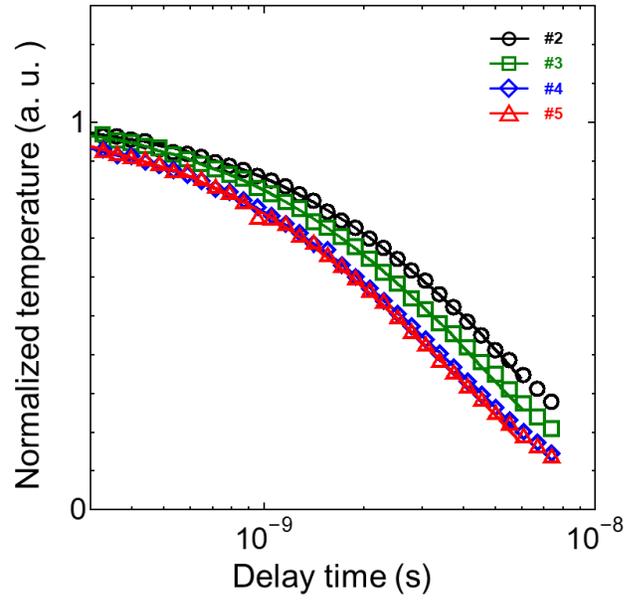

Fig. S3. Temperature decay profiles for the samples #2~#5 measured by TDTR, markers (squares, circles, spades, triangles) are experimental data, and curves correspond to the fitting results to the physical (heat conduction) model.

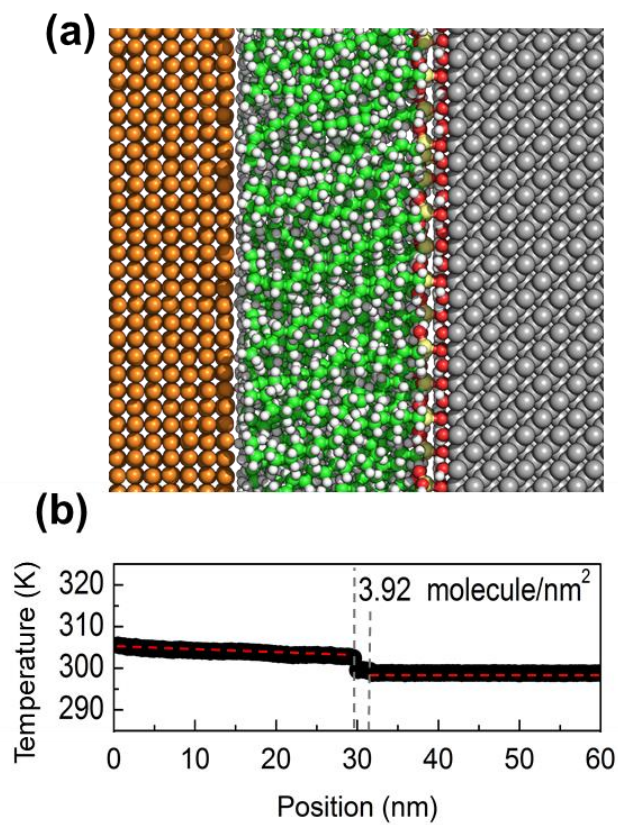

Figure S4. (a) Schematic of the interface and (b) corresponding temperature profile of a representative molecular dynamics simulation system (coverage 3.92 molecule/nm$^2$). The copper atoms are in orange and the carbon atoms of diamond are in gray. For SAMs, the silicon atoms are in yellow, oxygen atoms are in red, carbon atoms are in green, and hydrogen atoms are in white.

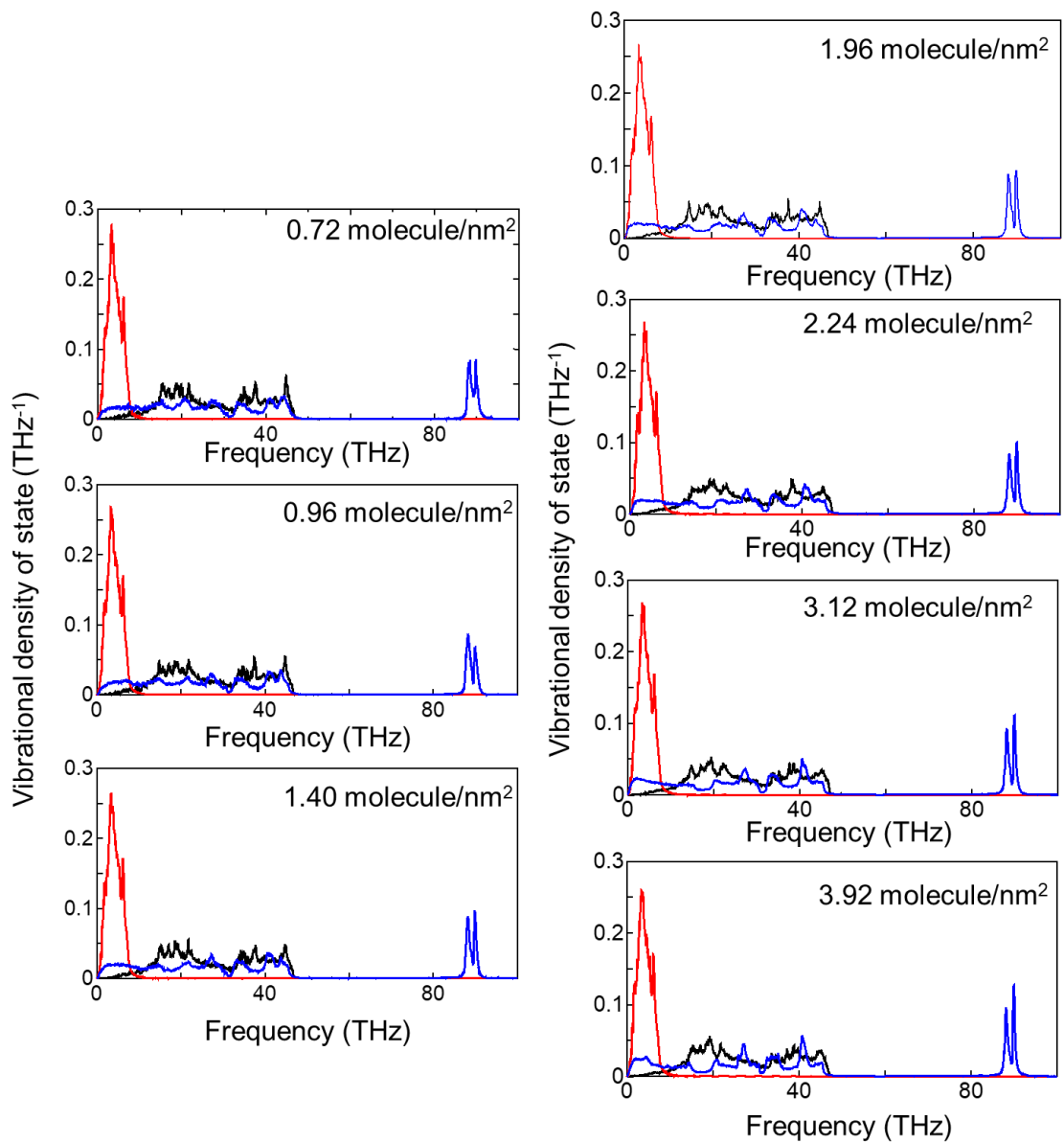

Figure S5. Vibrational density of states (vDOS) of copper (red curve), diamond (black curve), and SAM (blue curve) with different coverages.

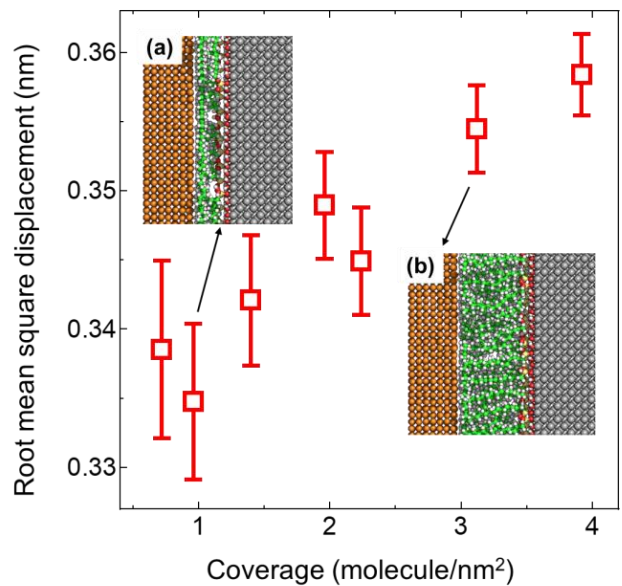

Figure S6. The root-mean-square displacement (RMSD) of SAM with different coverage. Insertions are two representative schematics of the system with (a) low (0.98 molecule/nm$^2$) and (b) high (3.12 molecule/nm$^2$) SAM coverages.

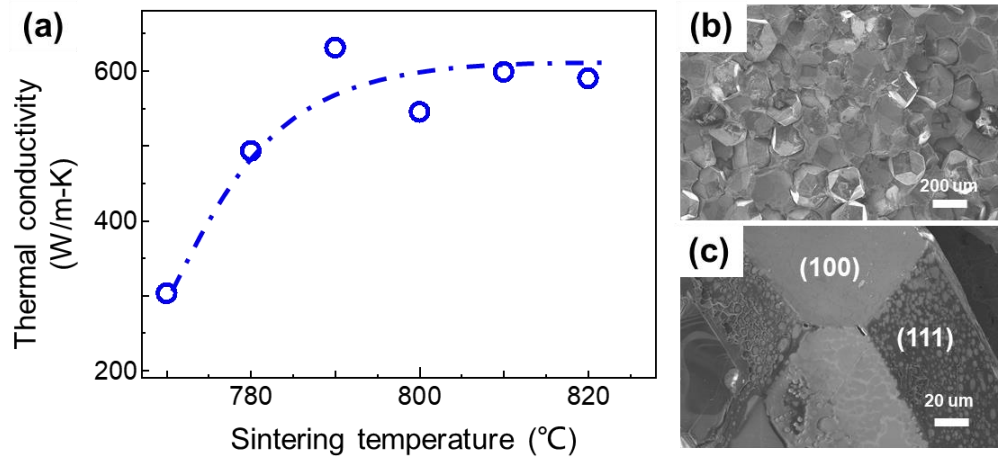

Figure S7. (a) Thermal conductivity of copper/diamond composite fabricated with pristine diamond under different sintering temperatures. (blue dash dot is the guide for the eye) (b)(c) SEM imagine of copper/diamond composite fabricated with pristine-diamond under 790 °C, with thermal conductivity of around 630 W/K-m.

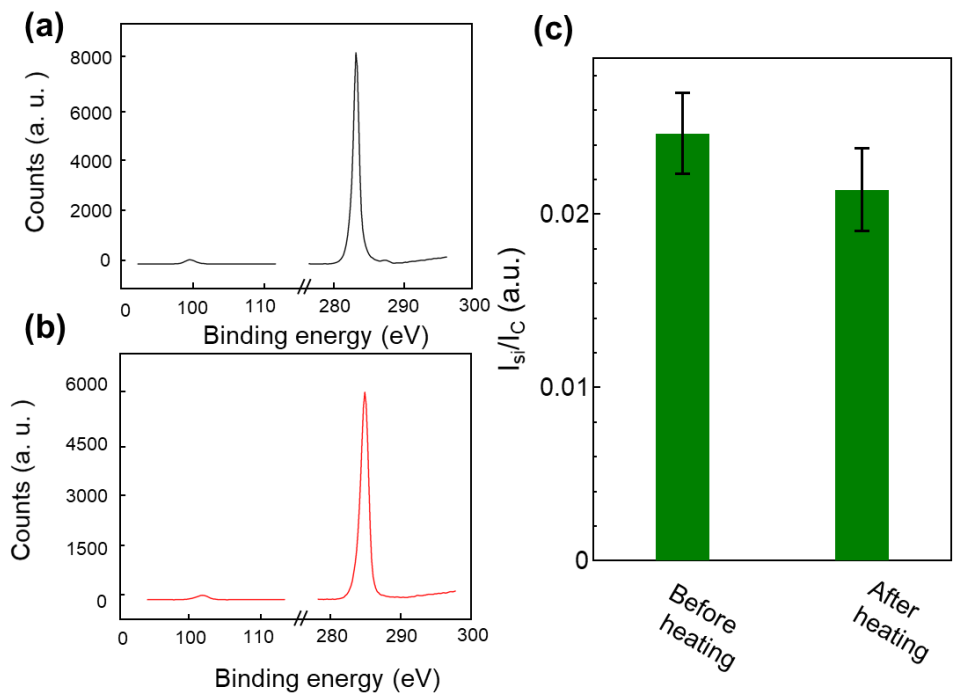

Figure S8. XPS spectrum (C 1s and Si 2p) of SAM-diamond powder (a) before and (b) after heating; (c) corresponding intensity ratio of Si 2p and C 1s XPS peaks ($I_{Si}/I_C$) for SAM-diamond powder.

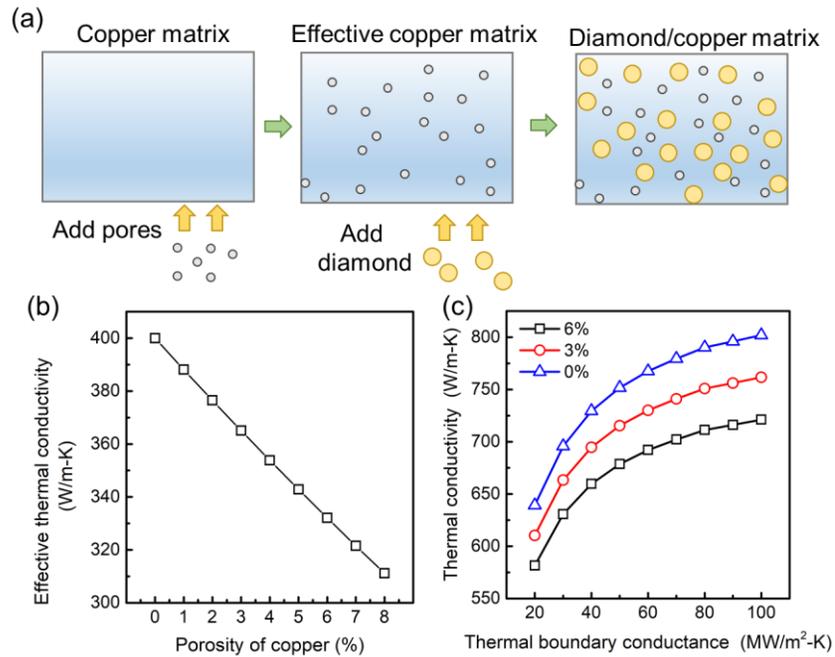

Figure S9. (a) Schematics of a two-step approach to estimate the thermal conductivity of copper/diamond composite with pores. (b) Effective thermal conductivity of copper with different porosities. (c) Thermal conductivity of the copper/diamond composite with different values of thermal boundary conductance (20~100 MW/m2-K) and porosity (0%, 3%, 6%).